\documentstyle[aps,preprint]{revtex}
\begin{document}
\draft{}
\title{
\begin{center}
 Phys. Rev. D 62, 058501 (2000)\\
\end{center}
 Comment on
 "Radiative decays, nonet symmetry and SU(3) breaking".}
\author{N.N. Achasov\thanks{E-mail address: achasov@math.nsc.ru}
and
 V.V. Gubin\thanks{E-mail address: gubin@math.nsc.ru}}
\address{Laboratory of Theoretical Physics\\
S.L. Sobolev Institute for Mathematics\\
630090 Novosibisk 90,\  Russia}
\date{\today}
\maketitle

\begin{abstract}
We comment on the paper by Benayoun, DelBuono, Eidelman,
Ivanchenko, and O'Connell [Phys. Rev. {\bf D 59}, 114027 (1999)].
 We show
that the decay $\phi\to\pi^0\gamma$ is absent in model of the
nonet symmetry and SU(3) breaking suggested by authors.
\end{abstract}
 A recent paper \cite{ben} reexamined "the problem
 of simultaneously describing in a consistent way all radiative and
 leptonic decays of light mesons
 ($V\to P\gamma$, $P\to V\gamma$, $P\to\gamma\gamma$, $V\to e^+e^-$)".
Unfortunately, this reexamination  cannot help but provoke
objections. The authors use the broken
$\mbox{U}_{\mbox{L}}(3)\times\mbox{U}_{\mbox{R}}(3)$ Lagrangian of
the hidden local symmetry approach \cite{hls} written in terms of
ideally mixed $\omega$ and $\phi$ states  in the following way:
\begin{eqnarray}
\label{hlsl}
&& L=...+\frac{1}{2}af_{\pi}^2g^2[(\rho^0_\mu)^2+(\omega^I_\mu)^2+
l_V(\phi^I_\mu)^2]-\\\nonumber
&& - aef_{\pi}^2g[\rho^0_\mu+\frac{1}{3}\omega^I_\mu+
l_V\frac{\sqrt{2}}{3}\phi^I_\mu]A_\mu +
\frac{1}{2}C\epsilon_{\mu\nu\rho\delta}
\partial_{\mu}\rho_{\nu}\partial_{\rho}\omega^I_{\delta}\pi^0+...\
\ .
\end{eqnarray}

 This expression  describes the ideally mixed vector mesons
[ $\omega^I=(u\bar u+d\bar d)/\sqrt{2}$, $\phi^I=-s\bar s$ ], the
kinetic and mass terms, their coupling to the electromagnetic
field, and  hadron states except for the coupling of the  $\phi$
meson to the $\rho\pi$ states and consequently to $\gamma\pi^0$
state.

The authors of Ref. \cite{ben} wanted to introduce this coupling
taking into account a deviation from the ideal mixing of the
$\omega$ and $\phi$ mesons. As is generally known, see, for
example, Ref.\cite{sakurai}, to do this in the tree approximation
( just that very approximation was used in Ref.\cite{ben} ) the
mixing terms should be introduced into the  mass and/or kinetic
terms of the Lagrangian, then the part of the Lagrangian quadratic
in fields should be diagonalized to get physical fields, and then
the part of the Lagrangian with interaction should be reexpressed
in terms of physical fields. It is the hard-and-fast rule, see
also, for example, the diagonalization of axial-vector and
pseudoscalar fields or the dynamical and external gauge fields in
the hidden local symmetry Lagrangian \cite{hls}.

Instead the authors of Ref.\cite{ben} reexpressed the Lagrangian
(\ref{hlsl}) in terms of "physical" (rotated) fields ($\omega_\mu$
and $\phi_\mu$) and got the following expression:
\begin{eqnarray}
\label{rothlsl}
L=&&\frac{1}{2}af_{\pi}^2g^2\left [(\rho^0_\mu)^2+
(\omega_\mu)^2\left (\cos^2\delta_V+l_V\sin^2\delta_V\right )+
(\phi_\mu)^2\left (\sin^2\delta_V+l_V\cos^2\delta_V\right )\right ]+\\
\nonumber
&&+(\phi_\mu\omega_\mu)af_{\pi}^2g^2(l_V-1)\sin\delta_V\cos\delta_V-\\
\nonumber
&&-aef_{\pi}^2g\left [\rho^0_\mu+\frac{1}{3}\omega_\mu\left (\cos\delta_V+
l_V\sqrt{2}\sin\delta_V\right )-\frac{1}{3}\phi_\mu\left (\sin\delta_V-
l_V\sqrt{2}\cos\delta_V\right )\right ]A_\mu +\\ \nonumber
&&+\frac{1}{2}C\cos\delta_V\epsilon^{\mu\nu\rho\delta}
\partial_{\mu}\rho_{\nu}\partial_{\rho}\omega_{\delta}\pi^0-
\frac{1}{2}C\sin\delta_V\epsilon_{\mu\nu\rho\delta}
\partial_{\mu}\rho_{\nu}\partial_{\rho}\phi_{\delta}\pi^0+...\ ,
\end{eqnarray}
where $\delta_V$ is an angle which describes the deviation
from the ideal mixing angle.

One can see that the last Lagrangian has the nondiagonal square
$\omega_\mu\phi_\mu$ term which describes the $\omega-\phi$
transitions and, hence, describes the nonphysical $\omega_\mu$ and
$\phi_\mu$ fields which do not have the definite masses. What
fields are physical is decided not by one author or another but by
the Lagrangian. In our case the $\omega^I_\mu$ and $\phi^I_\mu$
fields are physical.

The authors of Ref. \cite{ben} ignored this fact and got formulas
dependent on the $\delta_V$ angle:
\begin{eqnarray}
\label{formulae} && A(\phi\to\rho\pi\to\gamma\pi)=
-C\frac{e}{2g}\sin\delta_V\,,\\ \nonumber && A(\phi\to\gamma\to
e^+e^-)=aef_{\pi}^2g\frac{1}{3}\left (l_V\sqrt{2}\cos\delta_V
-\sin\delta_V\right )\frac{e}{m_{\phi}^2}\,, \nonumber
\end{eqnarray}
and so on. Hereafter we drop the obvious Lorentz structures.

Then they used these formulas and found the mixing parameters from
the data. The result is $\delta_V=-3.33 \pm 0.16^{\circ}$ and
$l_V=1.376\pm0.031$ which gives the $\phi$ meson mass equal to
$920\pm15$ MeV, see Eq.(1). But, the real trouble is that the
$\delta_V$ dependence is an artifact due to misuse of the
Lagrangian (\ref{rothlsl} ).

It is instructive to show this to the first order in  mixing
taking into account the nondiagonal $\omega-\phi$ term in the
Lagrangian (\ref{rothlsl} ):
\begin{eqnarray}
\label{show}
&& A(\phi\to\gamma\pi)= A(\phi\to\rho\pi\to\gamma\pi)+
 A(\phi\to\omega\to\rho\pi\to\gamma\pi)= -C\frac{e}{2g}\sin\delta_V+\\
\nonumber
&& +\frac{ea^2f_{\pi}^4g^3(l_V-1)\sin\delta_V\cos\delta_V C\cos\delta_V}
{2a^2f_{\pi}^4g^4\left (\sin^2\delta_V+l_V\cos^2\delta_V-\cos^2\delta_V-
l_V\sin^2\delta_V\right )}\simeq \\ \nonumber
&&-C\frac{e}{2g}\sin\delta_V+\frac{eC(l_V-1)\sin\delta_V}{2g(l_V-1)}=0\,,
\\[1pc] \nonumber
&& A(\phi\to e^+e^-)= A(\phi\to\gamma^*\to e^+e^-)+
A(\phi\to\omega\to\gamma^*\to e^+e^-)=\\  \nonumber
&& = aef_{\pi}^2g\frac{1}{3}\left (l_V\sqrt{2}\cos\delta_V-\sin\delta_V\right)
\frac{e}{m_{\phi}^2}+\\  \nonumber
&&+\frac{af_{\pi}^2g^2(l_V-1)\sin\delta_V\cos\delta_V}{af_{\pi}^2g^2\left
(\sin^2\delta_V+l_V\cos^2\delta_V-\cos^2\delta_V-l_V\sin^2\delta_V\right)
}aef_{\pi}^2g\frac{1}{3}\left (\cos\delta_V+l_V\sqrt{2}\sin\delta_V\right)
\frac{e}{m_{\phi}^2}\simeq\\ \nonumber
&&\simeq\frac{ae^2f_{\pi}^2g}{3m_{\phi}^2}(\sqrt{2}l_V-\sin\delta_V+
\frac{(l_V-1)\sin\delta_V}{(l_V-1)})=
\frac{\sqrt{2}l_Vae^2f_{\pi}^2g}{3m_{\phi}^2}\,,\\  \nonumber
&&\mbox{and so on.}
\end{eqnarray}

So, all amplitudes, obtained from the Lagrangian (\ref{rothlsl})
to first order in the $\omega-\phi$ mixing are equal to the ones
obtained from the Lagrangian (\ref{hlsl}). This is also true at
higher orders, but it is necessary to keep in mind that the
calculation at higher orders demands taking into account
corrections to the $\phi$ and $\omega$ wave functions. Certainly,
this conclusion is very natural because taking into account all
orders of the $\omega-\phi$ mixing involves nothing more than the
diagonalization of the Lagrangian (\ref{rothlsl}) which returns us
to the Lagrangian (\ref{hlsl}). Summing up we conclude that Ref.
\cite{ben} did not solve the problem of the radiative
$\phi\to\gamma\pi^0$ decay.

\end{document}